\documentclass[preprint,12pt,authoryear]{elsarticle}

\usepackage{amssymb}
\usepackage[version=3]{mhchem} % Formula subscripts using \ce{}

\usepackage{amsmath}
\usepackage{xcolor}
 \usepackage[normalem]{ulem} 

\journal{Carbon}

\begin{document}

\begin{frontmatter}



\title{Gas-phase chemistry of C$_3$H$_3^+$: a key building block for carbon mass growth in space} %% Article title


\author[1]{Corentin Rossi}

\author[2]{Jean-Christophe Loison}

\author[1,3]{Anne P. Rasmussen}

\author[1,4,5]{Roland Thissen} 

\author[4,6]{Nandana Pattathadathil}

\author[4,5]{Christian Alcaraz}

\author[1]{Ugo Jacovella\corref{cor1}}
\ead{ugo.jacovella@cnrs.fr} %% Author name

\affiliation[1]{organization={Universit\'e Paris-Saclay, CNRS, Institut des Sciences Mol\'eculaires d'Orsay},%Department and Organization
city={Orsay},
postcode={91405}, 
country={France}}

\affiliation[2]{organization={Université Bordeaux, CNRS, Bordeaux INP, ISM, UMR 5255},%Department and Organization
city={Talence},
postcode={F-33400}, 
country={France}}

\affiliation[3]{
  organization={Department of Physics and Astronomy},
  addressline={Ny Munkegade 120, Aarhus University},
  city={Aarhus C},
  postcode={8000},
  country={Denmark}
}

\affiliation[4]{organization={Universit\'e Paris-Saclay, CNRS, Institut de Chimie Physique, UMR8000},%Department and Organization
city={Orsay},
postcode={91405}, 
country={France}}

\affiliation[5]{organization={SOLEIL},%Department and Organization
addressline={L’Orme des Merisiers},
city={Saint-Aubin},
postcode={91192}, 
country={France}}

\affiliation[6]{
  organization={ Department of Physics, University of Trento},
  addressline={Via Sommarive 14},
  city={Trento},
  postcode={38123},
  country={Italy}
}

%% Abstract
\begin{abstract}

The gas-phase chemistry of \ce{C3H3+} plays a central role in the growth of carbonaceous species across a wide range of environments, from combustion systems to interstellar media. However, the presence of multiple isomers, primarily the linear propargyl cation and the cyclic cyclopropenyl cation, has so far prevented the determination of reliable isomer-specific reaction kinetics. In this work, we combine tunable vacuum-ultraviolet (VUV) photoionization with controlled ion–molecule reaction measurements to selectively produce and probe the reactivity of \ce{C3H3+} isomers.

By employing a titration approach, we extract isomer-specific reaction rate coefficients and branching ratios for reactions of \ce{C3H3+} with \ce{C2H4}, \ce{C3H4}, and \ce{C3H6}. Our results demonstrate a strong contrast in reactivity: the most stable cyclic isomer is found to be essentially unreactive under the investigated conditions, whereas the linear isomer exhibits efficient and diverse chemistry leading to the formation of larger hydrocarbon ions. In particular, the formation of \ce{C6H5+} is observed, providing an important intermediate in pathways toward aromatic species.

These findings help resolve long-standing ambiguities in previously reported rate coefficients and provide new quantitative constraints for kinetic and astrochemical models. More broadly, this work highlights the importance of isomer-specific chemistry in carbon growth processes and demonstrates the effectiveness of VUV-based photoionization methods for disentangling complex reaction networks in the gas phase.

\end{abstract}



%% Keywords
\begin{keyword}

Mass spectrometry \sep Ion-molecule reactions \sep Astrochemical modeling \sep Synchrotron radiation



\end{keyword}

\end{frontmatter}



\section{Introduction}

Understanding the molecular processes that govern early-stage carbon mass growth remains a central challenge in carbon science, spanning environments as diverse as combustion systems, plasma reactors, and astrophysical media. Among the key intermediates implicated in these processes, the \ce{C3H3+} hydrocarbon ion occupies a particularly important position. It acts as an efficient conduit for the conversion of small hydrocarbons into larger carbonaceous species \citep{smyth1982,herbst1983,baykut1986,loison2017,cherepanov2024,loison2026}, thus contributing to the formation of aromatic molecules, polycyclic aromatic hydrocarbons (PAHs), and ultimately carbonaceous particles.

In high-temperature environments such as flames and plasmas, neutral chemistry is generally expected to dominate, although ion chemistry can be non-negligible in certain cases \citep{smyth1982,baykut1986,cherepanov2024}. Beyond combustion, \ce{C3H3+} is well established as a key intermediate in astrochemical reaction networks that drive molecular complexity in low-temperature environments \citep{loison2017,loison2026}, including molecular clouds and planetary atmospheres. The high reactivity of \ce{C3H3+} towards small hydrocarbons, together with its central role in the formation of C$_3$H$_x$ neutrals and larger carbon chains, highlights its importance as a fundamental molecular building block in the pathways of growth of carbon under a wide range of physical conditions \citep{herbst2009,vuitton2019}.
A defining characteristic of \ce{C3H3+} is its existence in multiple isomeric forms, especially the linear propargyl cation (\ce{H2C3H+}/l-\ce{C3H3+}), which has been recently detected in the dense molecular cloud TMC1 by \cite{silva2023}, and the cyclic cyclopropenyl cation (c-\ce{C3H3+}), the latter being the most stable isomer. These isomers differ significantly in their geometry, electronic structure, and stability, leading to different spectroscopic signatures and reactivity patterns \citep{ausloos1981,smith1987,wiseman1990,mcewan1994,wyss2001,botschwina2011,huang2011}. The lowest isomerization barrier from propargylium to the cyclic isomer is of 2.1\,eV, as calculated by \cite{Maluendes1993a}. Both forms are expected to coexist in gas-phase environments, with their relative populations depending sensitively on formation pathways, internal energy distributions, and collisional conditions. As a result, carbon growth pathways involving \ce{C3H3+} are intrinsically isomer-dependent.

Despite its recognized importance, obtaining reliable reaction rate coefficients for \ce{C3H3+} remains a challenge \citep{ausloos1981,anicich1984,smith1987,ozturk1987,wiseman1990,mcewan1994,scott2000,savic2005}. One major difficulty arises from the simultaneous production of multiple isomers in laboratory sources, often without precise control over their relative abundances. Since the propargyl and cyclopropenyl cations may exhibit markedly different reaction cross sections and branching ratios, experimentally measured rate coefficients often represent convoluted averages rather than isomer-specific values. In general, \ce{C3H3+} reactivity shows a clear trend, which is that the cyclic isomer is largely unreactive, while the linear isomer displays rich and diverse chemistry toward neutral species. This limitation hinders the provision of accurate input data for astrochemical models and obscures mechanistic interpretation.

Additionally, in this study, we show that, by successively producing c-\ce{C3H3+} alone and then a mixture of cyclic and linear forms using vacuum-ultraviolet (VUV) dissociative photoionization, it is possible to perform gas-phase titration experiments. This approach relies on recent theoretical work by \cite{mallo2025}, in which the rate coefficient at 300~K was calculated for the reaction l-\ce{C3H3+} + \ce{C2H4}, and allows us to determine new isomer-specific reaction rate coefficients and refine previously reported values that were convoluted by isomeric mixtures.

\section{Experimental}
\label{Exp}

Parent ions (\ce{C3H3+}) were generated by dissociative photoionization of allene and propyne using tunable radiation in the 10.5–15 eV photon energy range at the DESIRS undulator-based vacuum ultraviolet (VUV) beamline \citep{nahon2012} of the SOLEIL synchrotron facility in Saint-Aubin, France. Higher-order radiation was suppressed by means of a gas filter filled with Kr.

Ion–molecule reactions were investigated using the CERISES apparatus, which consists of a tandem mass spectrometer (quadrupole–octopole–octopole–quadrupole) coupled to a photoionization source \citep{CunhaDeMiranda2015, rossi_tunable_2023}. The \ce{C3H3+} parent ions produced in the source were first mass-selected by the initial quadrupole and subsequently guided through an octopole into the reaction cell, where the collision energy could be precisely controlled. Reaction product ions were then radially confined, extracted, and analyzed by the second quadrupole mass filter.

Absolute reaction cross sections were determined from measurements of the neutral reactant pressure in the reaction cell together with the recorded intensities of parent and product ions. Measurements were performed over the full photon energy range while maintaining a fixed center-of-mass collision energy of 0.3\,eV. Neutral reactants were introduced at low pressures (typically below 80\,nbar) to ensure that attenuation of the parent ion signal remained below 10\,\%, thereby preserving single-collision conditions. Reaction rate coefficients were derived from the measured absolute cross sections following the procedure described by \cite{rossi2025}.

\section{Results and discussion}

\subsection{Reactivity of \ce{C3H3+} with \ce{C2H4}}

\begin{figure}[!h]
\includegraphics[width=\linewidth]{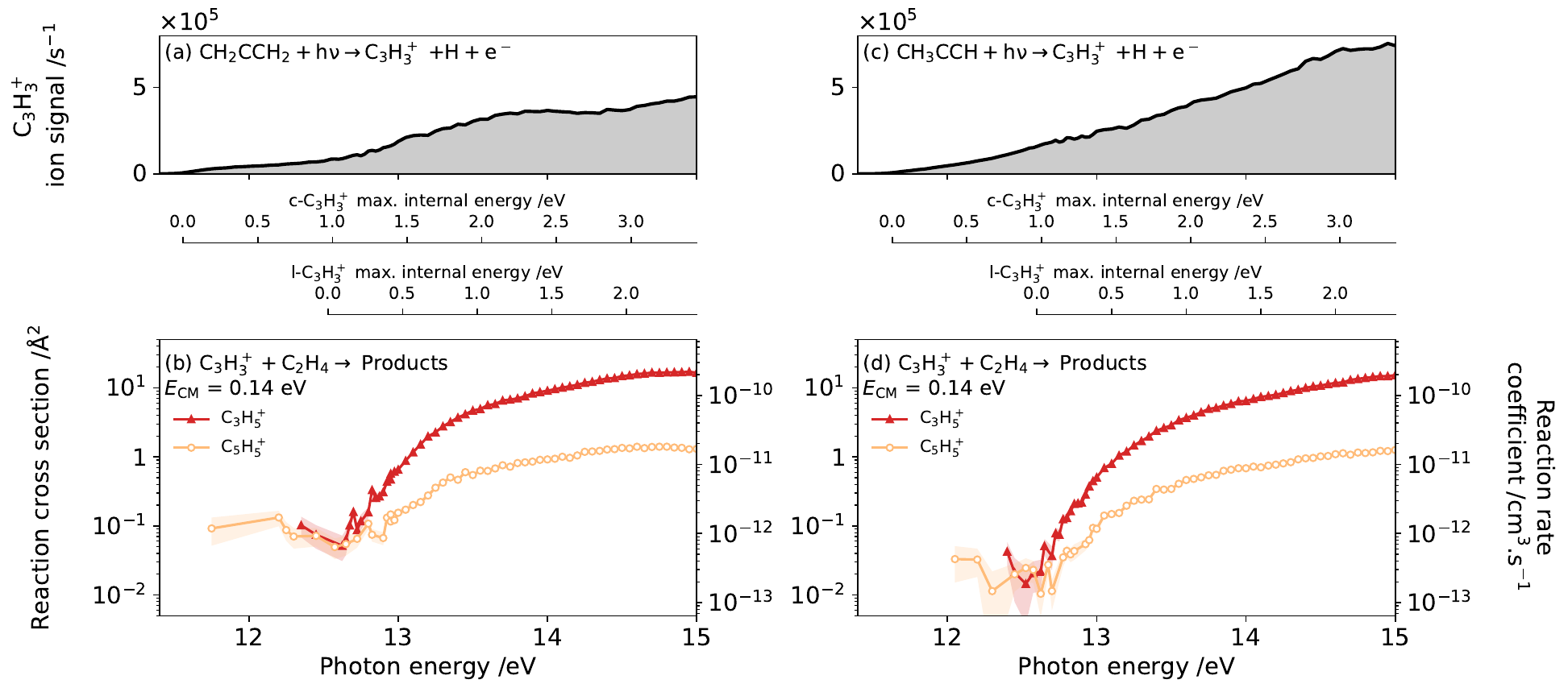}
\caption{Panels a and c display the \ce{C3H3+} ion signal from the dissociative photoionization of allene and propyne, respectively. Panels b and d show the reaction cross sections (left axis) and rate coefficients (right axis) with \ce{C2H4}, as a function of photon energy. The secondary $x$-axes indicate the maximum internal energy deposited in the system for the cyclic (top axis) and the linear (bottom axis) of \ce{C3H3+}.} 
\label{FigC2H4}
\end{figure}

Panels b and d in Figure~\ref{FigC2H4} present the reaction cross sections (left axis) and rate coefficients (right axis) for reactions of \ce{C3H3+} with \ce{C2H4} as a function of photon energy, where \ce{C3H3+} are generated from allene and propyne, respectively. Only reaction channels that exceed 1 \AA$^2$ at 15 eV are shown (same apply for other reactions studied in the realm of this paper). The maximum internal energies of the \ce{C3H3+} ions are given at the top of panels b and d and are defined as the photon energies minus the appearance energies (AE) of the cyclic and linear isomers, as calculated by \cite{mebel2008}.

As expected, the propargyl structure is significantly more reactive than the cyclic form. This is supported by the weak reactivity (< 0.1 \AA$^2$) observed below the AE of propargyl. However, the flat reaction cross section in the 11.8-12.5 eV range, where only c-\ce{C3H3+} is produced, is puzzling and raises questions regarding the reaction mechanism, which could potentially be explained by tunneling effects. Nevertheless, the reactivity of c-\ce{C3H3+} is too low to play a significant role in the growth of larger carbonaceous species and will therefore not be discussed further in the present study.

The two reaction channels correspond to the formation of \ce{C3H5+} and \ce{C5H5+}(loss of \ce{C2H2} or \ce{H2} from the adduct intermediate), with an approximately equal branching ratio at the AE threshold, which can be correlated to a temperature of 300 K. As the internal energy of the propargyl ions increases, the \ce{C3H5+} channel becomes more dominant, consistent with the predictions of \cite{mallo2025}.

Previously, the only reported measurements were those of \cite{smyth1982}, in which both isomers were formed in an uncontrolled manner by electron ionization of allene, propyne, or propargyl chloride. In that experiment, two reaction channels leading to \ce{C5H7+} and \ce{C5H5+} were observed, with a total rate coefficient of \(1.1 \times 10^{-9}\) cm$^{3}$ s$^{-1}$ at 300 K. This value is in good agreement with the calculations of \cite{mallo2025} for the total reaction rate, obtained as the sum of the two predicted channels, \ce{C5H5+} and \ce{C3H5+}. Yet, our results confirm the discrepancy in the reported product channels highlighted by \cite{mallo2025}. Although we cannot provide an explanation for this discrepancy in the observed channels, we independently validate the theoretical results reported by \cite{mallo2025}.

\begin{figure}[!h]
\includegraphics[width=\linewidth]{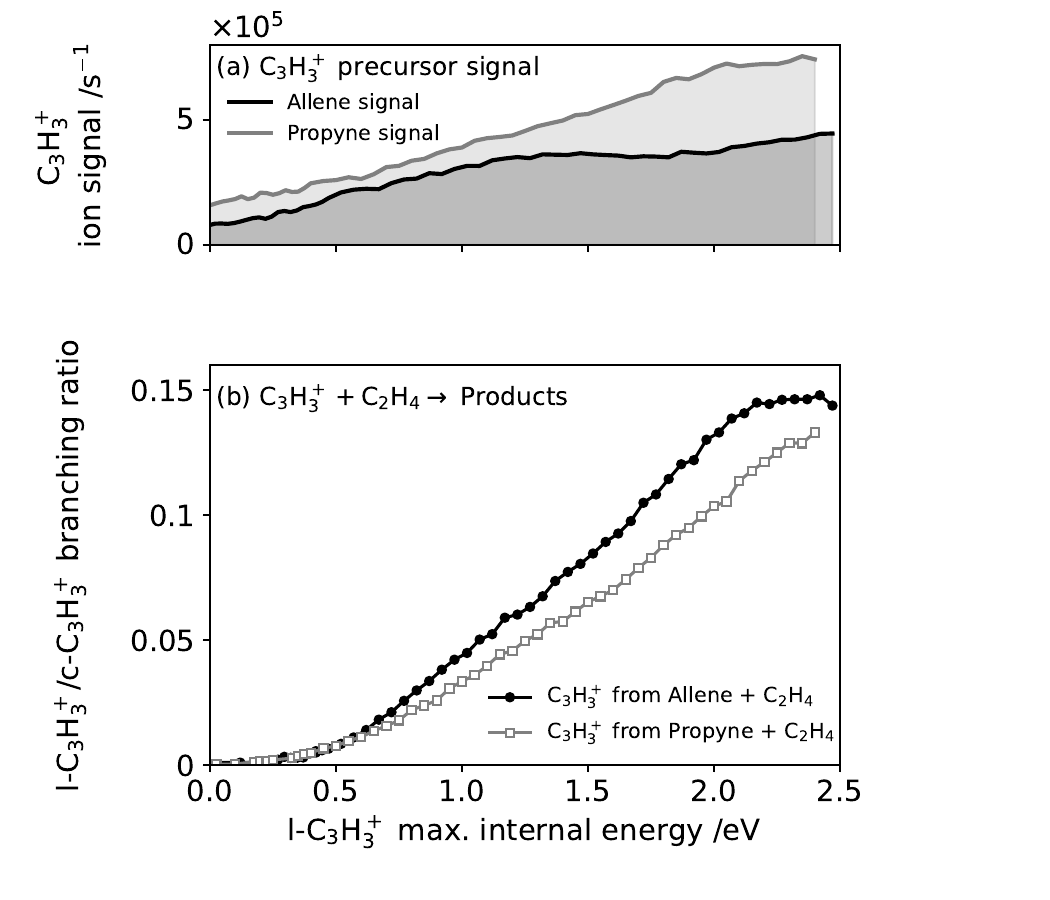}
\caption{Panels a displays the \ce{C3H3+} ion signal from the dissociative photoionization of allene and propyne, respectively. Panels b shows the sum of rate coefficient of the two main reaction product of \ce{C3H3+} + \ce{C2H4}, as a function of the maximum internal energy deposited in the system for the linear \ce{C3H3+}.} 
\label{FigC2H4_BR}
\end{figure}

While this reaction of \ce{C3H3+} + \ce{C2H4} does not play a significant role in the route towards five- or six-membered ring molecules (\cite{mallo2025,jacovella2026,loison2026}), it can serve two purposes. (i) It shows that the proportion of cyclic versus propargyl species produced through dissociative photoionization of allene and propyne is the same, indicating that both precursors can be used interchangeably. (ii) This reaction can be used to titrate the abundance of the propargyl form, as illustrated in Figure~\ref{FigC2H4_BR}. The figure displays the sum of the reaction rates of the two reported reaction channels, as a function of the maximum internal energy deposited in the l-\ce{C3H3+} isomer, relative to the value reported by \cite{mallo2025} ($1.6 \times 10^{-9}$~cm$^3$.s$^{-1}$). It shows that up to 15\% of propargylium can be generated at 15\,eV. Three assumptions had to be made to obtain this result. First, it was assumed that the reaction was temperature-independent, which is a reasonable approximation given the low values of the parameters $\beta$ calculated by \cite{mallo2025}. Second, it was assumed that no isomerization from the linear to the cyclic structure occurs, or only very little, upon collision with the neutral reaction partner (which is also an approximation performed by \cite{mallo2025}). Third, the cyclic isomer exhibits low reactivity even at the highest photon energies. This may either reflect the presence of a high activation barrier or indicate that, despite the high photon energies, the cyclic isomer retains only a limited amount of internal energy, with most of the excess energy being shared between the photoelectron energy and the kinetic energy of the hydrogen atom during the dissociation of \ce{C3H4+}.

\subsection{Reactivity of \ce{C3H3+} with \ce{C3H4} (allene and propyne)}

\begin{figure}[!h]
\includegraphics[width=\linewidth]{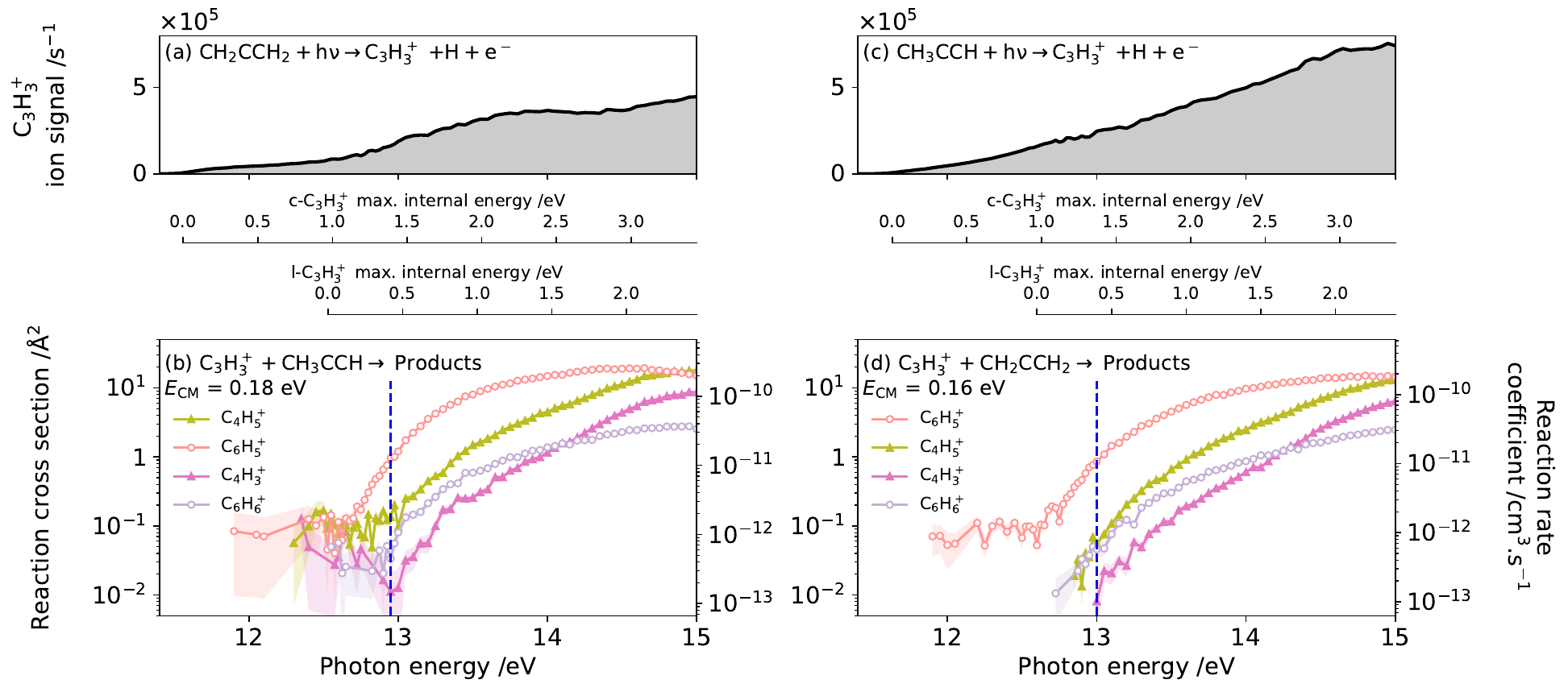}
\caption{Panels a and c display the \ce{C3H3+} ion signal from the dissociative photoionization of allene and propyne, respectively. Panels b and d show the reaction cross sections (left axis) and rate coefficients (right axis) with propyne and allene, respectively, as a function of photon energy. The secondary $x$-axes indicate the maximum internal energy deposited in the system for the cyclic (top axis) and the linear (bottom axis) of \ce{C3H3+}. The dashed-blue line marks the value at which values are reported in tables~\ref{tab_allene} and \ref{tab_propyne}.} 
\label{FigC3H4}
\end{figure}

Panels b and d in Figure~\ref{FigC3H4} present the reaction cross sections (left axis) and rate coefficients (right axis) for reactions of \ce{C3H3+} with propyne and allene, respectively, as a function of photon energy. Only reaction channels that exceed 1~\AA$^2$ at 15~eV are shown (figures showing all followed reaction channels are included in the Supplementary Information). 

A previous experimental study by \cite{anicich1984} reported no chemical reactivity of \ce{C3H3+} with allene or propyne, as the \ce{C3H3+} ions were produced by electron ionization, which predominantly leads to the formation of the unreactive c-\ce{C3H3+} isomer. \cite{anicich1984} mentioned the reactivity of propargylium but did not provide the experimental conditions or reaction rate coefficients, only branching ratios were reported. These results were stated to originate from \cite{lehman1976}, however, we were unable to retrieve this information from the cited reference.

\begin{table}[h]
\centering
\caption{Reaction rates and branching ratios for l-\ce{C3H3+} + propyne. Values are taken at a maximum internal energy deposited in l-\ce{C3H3+} of 0.4~eV (which corresponds to a photon energy of 12.9~eV.)}
\begin{tabular}{l c cc}
\hline
\textbf{Product} & \textbf{Reaction rate / cm$^{3}$ s$^{-1}$} & \multicolumn{2}{c}{\textbf{Branching ratio}} \\
\cline{3-4}
 &  & \textbf{This work} & \textbf{\cite{anicich1984}} \\
\hline

\ce{C6H5+} & $2.0 \times 10^{-9}$  & 0.81 & 0.45 \\
\ce{C4H5+} & $3.2 \times 10^{-10}$ & 0.13 & 0.35 \\
\ce{C6H6+} & $1.4 \times 10^{-10}$ & 0.06 & NA \\
\ce{C4H3+} & $1 \times 10^{-11}$ & >0.01 & 0.20 \\
\hline
\end{tabular}
\label{tab_propyne}
\end{table}

\begin{table}[h]
\centering
\caption{Reaction rates and branching ratios for l-\ce{C3H3+} + allene. Values are taken at a maximum internal energy deposited in l-\ce{C3H3+} of 0.4~eV (which corresponds to a photon energy of 13.0~eV.)}
\begin{tabular}{l c cc}
\hline
\textbf{Product} & \textbf{Reaction rate / cm$^{3}$ s$^{-1}$} & \multicolumn{2}{c}{\textbf{Branching ratio}} \\
\cline{3-4}
 &  & \textbf{This work} & \textbf{\cite{anicich1984}} \\
\hline

\ce{C6H5+} & $2.1 \times 10^{-9}$  & 0.89 & 0.38 \\
\ce{C4H5+} & $1.4 \times 10^{-10}$ & 0.06 & 0.37 \\
\ce{C6H6+} & $1.2 \times 10^{-10}$ & 0.05 & NA \\
\ce{C4H3+} & $2.0 \times 10^{-11}$ & >0.01 & 0.25 \\
\hline
\end{tabular}
\label{tab_allene}
\end{table}

Four reaction channels are identified, ordered by decreasing branching ratio at the propargylium AE: \ce{C6H5+}, \ce{C4H5+}, \ce{C6H6+}, and \ce{C4H3+}. Their branching ratios at a maximum internal energy deposition of 0.4~eV in l-\ce{C3H3+}, together with the corresponding propargylium reaction rate coefficients derived by titration with \ce{C2H4}, are reported in Tables \ref{tab_propyne} and \ref{tab_allene}. The reported reaction coefficients are measured at the lowest internal energy that yielded a detectable signal in all reaction channels, allowing the branching ratios to be determined.

The \ce{C6H5+} and \ce{C6H6+} profiles are similar, as are those of \ce{C4H5+} and \ce{C4H3+}, suggesting comparable reaction mechanisms within each pair. Notably, the formation of \ce{C6H5+} from propyne provides the missing link in the proposed benzene-formation pathway in dense molecular clouds of \cite{loison2026}; the corresponding branching ratios and rate coefficients are given in Tables \ref{tab_propyne} and \ref{tab_allene}.

\begin{figure}[!h]
\includegraphics[width=\linewidth]{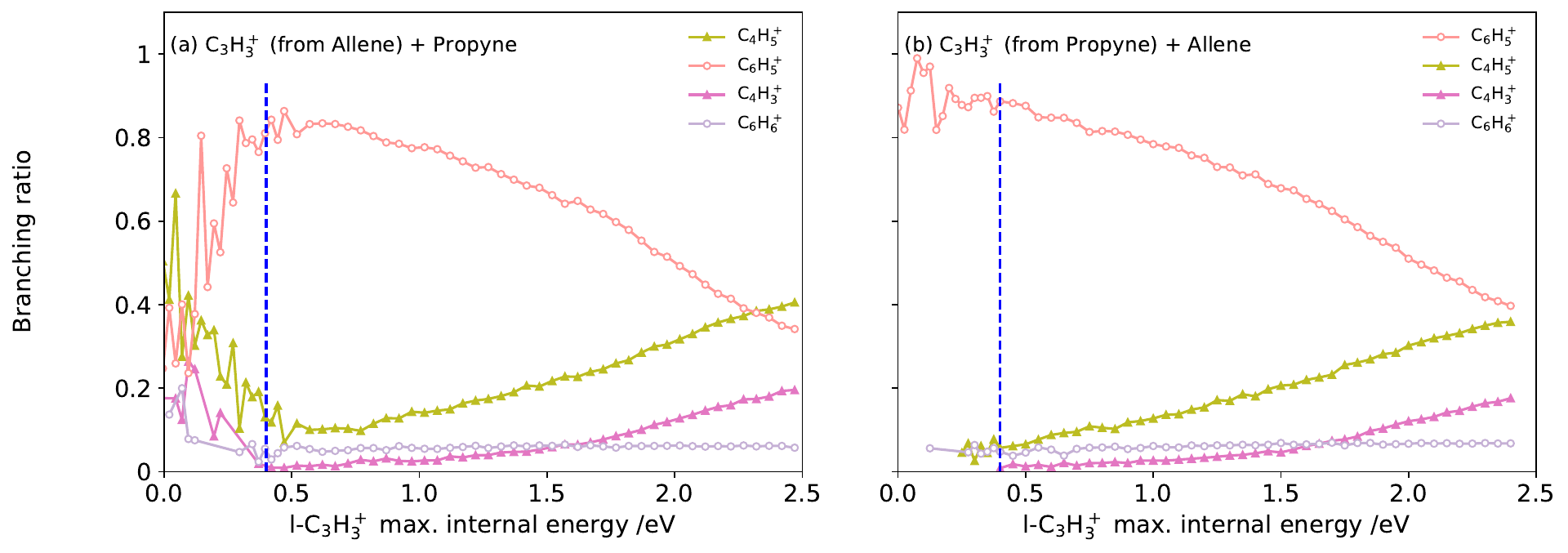}
\caption{Panels a and b depict the branching ratios as a function of the maximum internal energy for the reaction of l-\ce{C3H3+} with propyne and allene, respectively. The dashed-blue line marks the value at which values are reported in tables~\ref{tab_allene} and \ref{tab_propyne}.} 
\label{FigC3H4_BR}

\end{figure}

Moreover, Figure \ref{FigC3H4_BR} presents the evolution of the branching ratios as a function of internal energy. The channel leading to \ce{C6H5+} is favored at low internal energies, which are the most representative conditions in dense molecular clouds such as TMC-1.

\subsection{Reactivity of \ce{C3H3+} with \ce{C3H6}}

\begin{table}[h]
\centering
\caption{Reaction rates and branching ratios for l-\ce{C3H3+} + \ce{C3H6}. Values are taken at a maximum internal energy deposited in l-\ce{C3H3+} of 0.4~eV (which corresponds to a photon energy of 12.9~eV.)}
\begin{tabular}{l c c}
\hline
\textbf{Product} & \textbf{Reaction rate / cm$^{3}$ s$^{-1}$} & \textbf{Branching ratio} \\
\hline

\ce{C3H5+} & $5.5 \times 10^{-10}$  & 0.40 \\
\ce{C4H5+} & $4.6 \times 10^{-10}$ & 0.34 \\
\ce{C4H7+} & $2.0 \times 10^{-10}$ & 0.15 \\
\ce{C6H5+} & $7.6 \times 10^{-11}$ & 0.06 \\
\ce{C5H6+} & $6.3 \times 10^{-11}$ & 0.05 \\
\hline
\end{tabular}
\label{tab_c3h6}
\end{table}

\begin{figure}[!h]
\includegraphics[width=\linewidth]{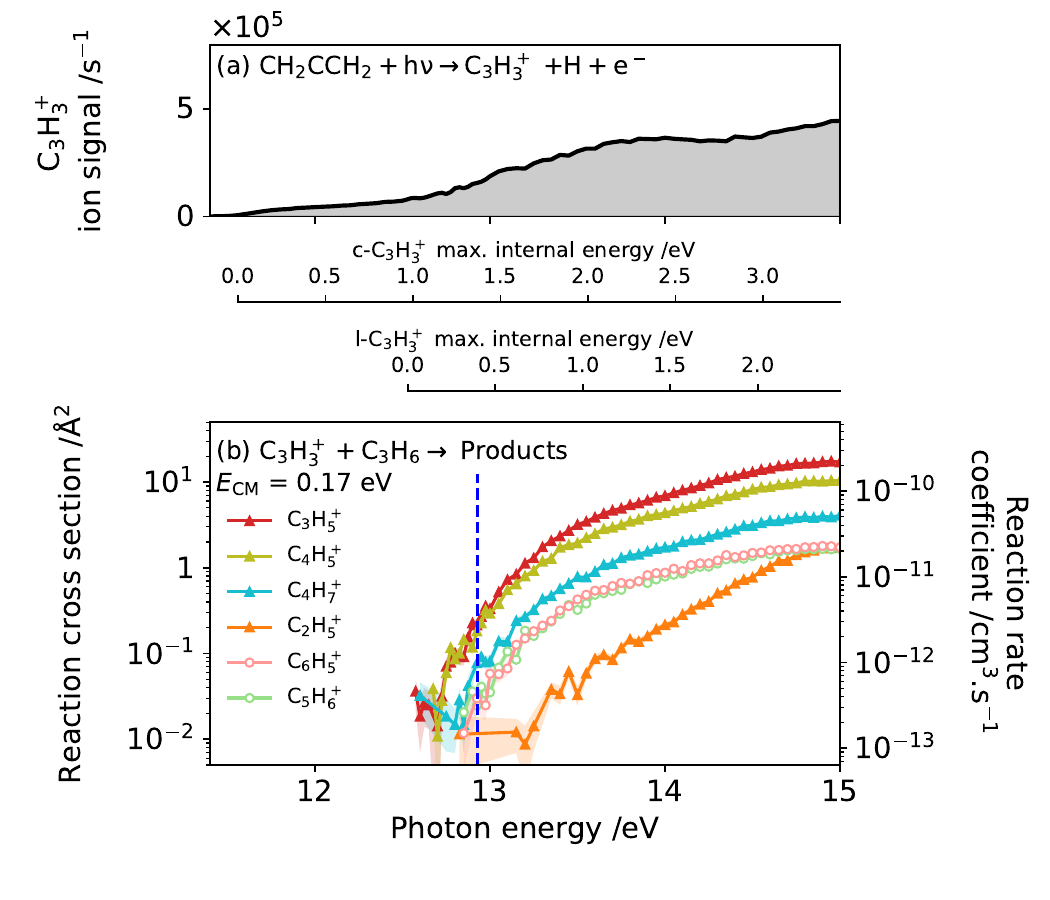}
\caption{Panel a displays the \ce{C3H3+} ion signal from the dissociative photoionization of allene. Panels b shows the reaction cross sections (left axis) and rate coefficients (right axis) with \ce{C3H6}, as a function of photon energy. The secondary $x$-axes indicate the maximum internal energy deposited in the system for the cyclic (top axis) and the linear (bottom axis) of \ce{C3H3+}. The dashed-blue line marks the value at which values are reported in table~\ref{tab_c3h6}} 
\label{FigC3H6}
\end{figure}

In the literature, the only reported study of this reaction concerns exclusively the channel leading to \ce{C4H5+} + \ce{C2H4}, with a reaction cross section ranging from 15.5 to 18.6~\AA$^2$, depending on the collision energy \citep{harrison1963}. Figure~\ref{FigC3H6} presents the reactivity of \ce{C3H3+}, generated from propyne, with \ce{C3H6}. In contrast to previous experimental work, six reaction channels are identified here and listed in Table~\ref{tab_c3h6}. All channels, except for the one leading to \ce{C2H5+}, exhibit similar reaction profiles, suggesting that they proceed via closely related underlying mechanisms. By contrast, the pathway leading to \ce{C2H5+} is likely governed by a distinct reaction mechanism.

\begin{figure}[!h]
\includegraphics[width=\linewidth]{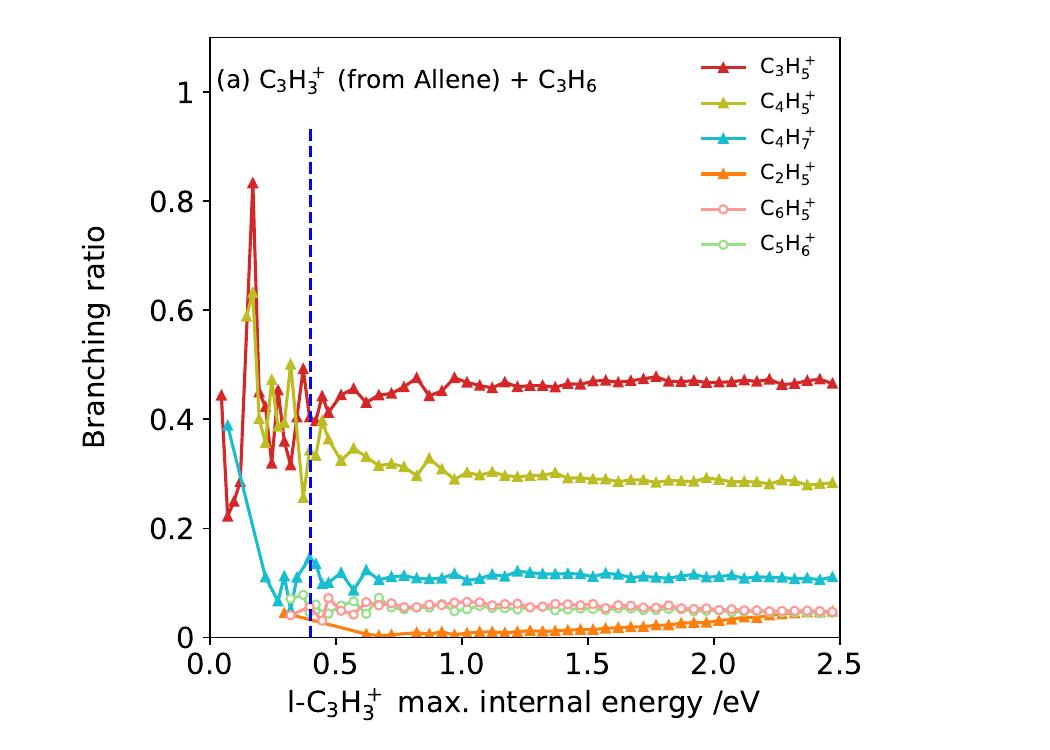}
\caption{Branching ratios as a function of the maximum internal energy for the reaction of l-\ce{C3H3+} with \ce{C3H6}. The dashed-blue line marks the value at which values are reported in table~\ref{tab_c3h6}.} 

\label{FigC3H6_BR}
\end{figure}

This is confirmed in Figure \ref{FigC3H6_BR}, where all the branching ratios except \ce{C2H5+} remain essentially constant, as expected if all channels proceed through the same adduct formation and share a common rate-limiting step.

\subsection{Astrochemical implications}

\begin{figure}[!h]
\includegraphics[width=\linewidth]{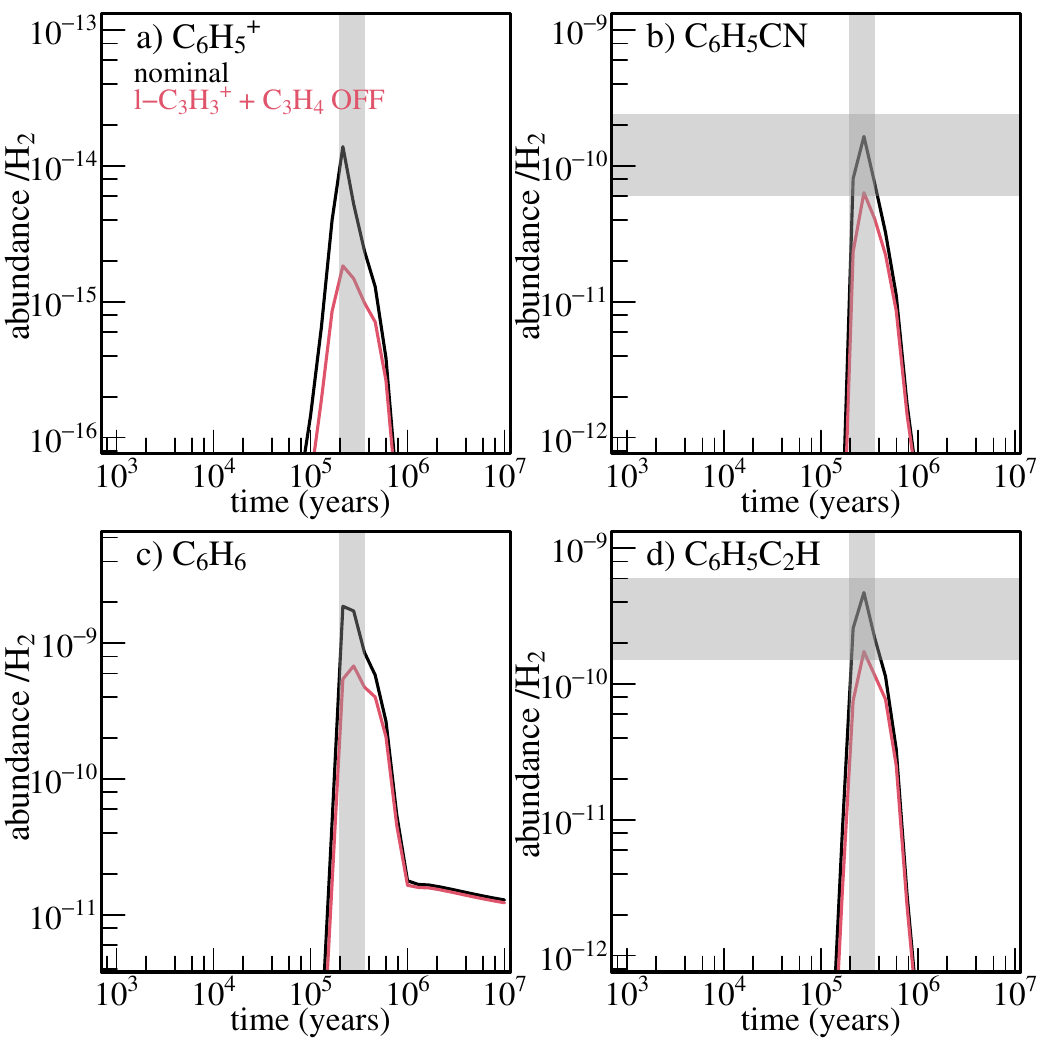}
\caption{Gas–grain astrochemical model results for the first aromatic species in dark clouds as a function of cloud age. Solid black line: updated network results. Solid red line: nominal network excluding the gas-phase l-\ce{C3H3+} + \ce{C3H4} reaction.  The vertical shaded rectangle indicates the most plausible chemical age given by the best agreement between the model predictions and the observed abundances of 77 key species. The horizontal shaded rectangles indicate the observed abundances in TMC-1 with an arbitrary error of $\pm 3$: \ce{C6H5CN} \citep{mcguire2018}, \ce{C6H5C2H} \citep{loru2023}} 
\label{Figab}
\end{figure}

To evaluate the role of the l-\ce{C3H3+} + \ce{C2H4}, \ce{C3H4}, and \ce{C3H6} reactions in the interstellar medium, we use the three-phase, time-dependent astrochemical model Nautilus with a recently updated chemical network. Nautilus treats the gas-phase, grain-surface, and grain-mantle chemistry simultaneously. The underlying chemical network is based on kida.uva.2024 \citep{wakelam2024} and has recently been extended to provide a more comprehensive description of aromatic chemistry in both the gas phase and on interstellar dust grains \citep{jacovella2026, loison2026}. A detailed description of the model and the updated chemical network can be found in our recent publications \citep{Loison2026a,jacovella2026, loison2026}.

Among the three studied reactions, the l-\ce{C3H3+} + \ce{C3H4} reaction is the most significant, as it initiates a reaction sequence  (followed by \ce{C6H5+} + \ce{H2} and then \ce{C6H7+} + \ce{e-}) that constitutes the dominant pathway towards benzene formation in Nautilus. In Figure~\ref{Figab}, we compare the nominal model predictions with the observational abundances of \ce{C6H5CN}, and \ce{C6H5C2H}. The model reproduces well the observed abundance of \ce{C6H5CN} and slightly overestimates that of \ce{C6H5C2H}. The differences compared with the results presented in our recent paper \citep{loison2026} arise from the use of a slightly different initial atomic oxygen and carbon abundances and from a more comprehensive treatment of the C + \ce{C6H6} reaction \citep{loison2026b, Hickson2026}. We also show in red the results from a model in which the l-\ce{C3H3+} + \ce{C3H4} reaction is omitted. Removing this reaction leads to a substantial decrease in aromatic abundances, by approximately a factor of 2.5. This reduction indicates that the formation of \ce{C6H5+} from l-\ce{C3H3+} + \ce{C3H4} reaction accounts for about 60\% of the \ce{C6H6} production, highlighting its key role in aromatic formation.

As noted in \cite{loison2026}, the reaction (c,l)$-$\ce{C3H3+} + \ce{H2} plays a crucial role in the formation of the smallest \ce{C6} aromatic compounds. This reaction presents an entrance barrier \citep{lin2013}, which limits the formation of \ce{C3H5+}, the main gas-phase precursor of \ce{C3H4}. Alternative formation pathways, such as \ce{CH} + \ce{C2H4}, contribute only minor fluxes. Grain-surface chemistry provides a non-negligible source of \ce{C3H4}; however, it remains insufficient to reproduce the observed abundances \citep{hickson2016}. Since the l-\ce{C3H3+} + \ce{C3H4} reaction is the major pathway for benzene production and many of the minor pathways involve other neutral or ionic \ce{C3H_$x$^{(+)}} species, standard astrochemical models that fail to reproduce the observed abundances of \ce{C3H_$x$^{(+)}} — particularly \ce{CH3CCH} — significantly underestimate the abundances of aromatic compounds. As described in \cite{loison2026}, to reproduce the observed abundances of \ce{CH3CCH}, we choose to use, without any theoretical calculations, low but non-negligible rates for l-\ce{C3H3+} + \ce{H2} that implicitly invokes the tunneling effect, consistent with the 300~K upper limits and the calculated energy barriers \citep{savic2005,lin2013}, allowing the reproduction of all \ce{C3H_$x$^{(+)}} observations, particularly \ce{CH3CCH} and l-\ce{C3H3+}.

The prevalent route involving l-\ce{C3H3+} in benzene formation can be attributed to several factors. First, l-\ce{C3H3+} is efficiently produced through the l-\ce{C3H+ + H2} reaction \citep{smith1987}. Second, its dissociative electron recombination is relatively slow (seven times slower than that of the cyclic isomer) \citep{McLain2005}, resulting in a comparatively high abundance for a molecular ion \citep{silva2023}. Finally, l-\ce{C3H3+} exhibits high reactivity toward unsaturated species, enabling efficient growth of carbon chains. Nevertheless, despite its high reactivity, the l-\ce{C3H3+} + \ce{C2H4} and l-\ce{C3H3+} + \ce{C3H6} reactions are much less prevalent in interstellar chemistry. This is because they involve lower reaction fluxes than the l-\ce{C3H3+} + \ce{C3H4} pathway, and the products they form can be produced more efficiently through alternative reaction pathways.

\section{Conclusions}

This work provides new insights into the gas-phase reactivity of \ce{C3H3+}, a key intermediate in carbon growth processes, by disentangling the contributions of its cyclic and linear isomers. Using tunable VUV photoionization combined with controlled ion–molecule reaction measurements, we demonstrate that isomer-selective production and titration approaches enable the determination of reliable isomer-specific reaction rate coefficients, overcoming a long-standing limitation associated with mixed isomer populations.

Our results confirm the strongly contrasted reactivity of the two isomers: while c-\ce{C3H3+} is essentially unreactive under the explored conditions, l-\ce{C3H3+} exhibits a rich and efficient chemistry with unsaturated hydrocarbons. Reactions with \ce{C2H4}, \ce{C3H4}, and \ce{C3H6} reveal multiple reaction pathways leading to larger hydrocarbon ions, including \ce{C6H5+}, thereby highlighting its role as a key intermediate in carbon mass growth and in the formation of aromatic precursors.

The isomer-specific rate coefficients and branching ratios reported here provide more robust constraints for kinetic and astrochemical models, particularly in environments where ion–molecule reactions govern molecular complexity. In addition, the identification of efficient pathways toward \ce{C6H5+} supports proposed mechanisms for benzene formation in interstellar environments.

Overall, this study emphasizes the critical importance of accounting for isomer-specific chemistry in the description of carbon growth processes and demonstrates the capability of VUV-based approaches to probe complex ion–molecule reaction networks with high selectivity and quantitative accuracy.

%\appendix
%\section{Example Appendix Section}
%\label{app1}

\section*{Acknowledgements}

This work was performed on the DESIRS Beamline at SOLEIL synchrotron under Proposal No. 99250229. We are grateful to the whole staff of SOLEIL for running the facility. The work was funded by the French “Agence Nationale de la Recherche” (ANR) under grant no. ANR-22-CE29-0013 (Project iSELECTION). This work was also supported by the Programme National ”Physique et Chimie du Milieu Interstellaire” (PCMI) and a research grant (VIL71404) from VILLUM FONDEN.

\section*{Data Availability}

The data supporting the findings of this study are openly available on Zenodo at 10.5281/zenodo.22650142.

\bibliographystyle{elsarticle-harv} 
\bibliography{C3H3+}

\newpage

%-------------------------------------------------------------------------

\begin{center}
{\LARGE Supporting information: Gas-phase chemistry of C$_3$H$_3^+$: a key building block for carbon mass growth in space
}\\ 
\end{center}

\vspace{0.5cm}

Corentin Rossi, Jean-Christophe Loison, Anne P. Rasmussen, Roland Thissen, Nandana Pattathadathil, Christian Alcaraz and Ugo Jacovella

\vspace{1cm}

\section*{Reactivity of \ce{C3H3+}}
\label{si:reac}

\begin{figure}[!h]
\includegraphics[width=\linewidth]{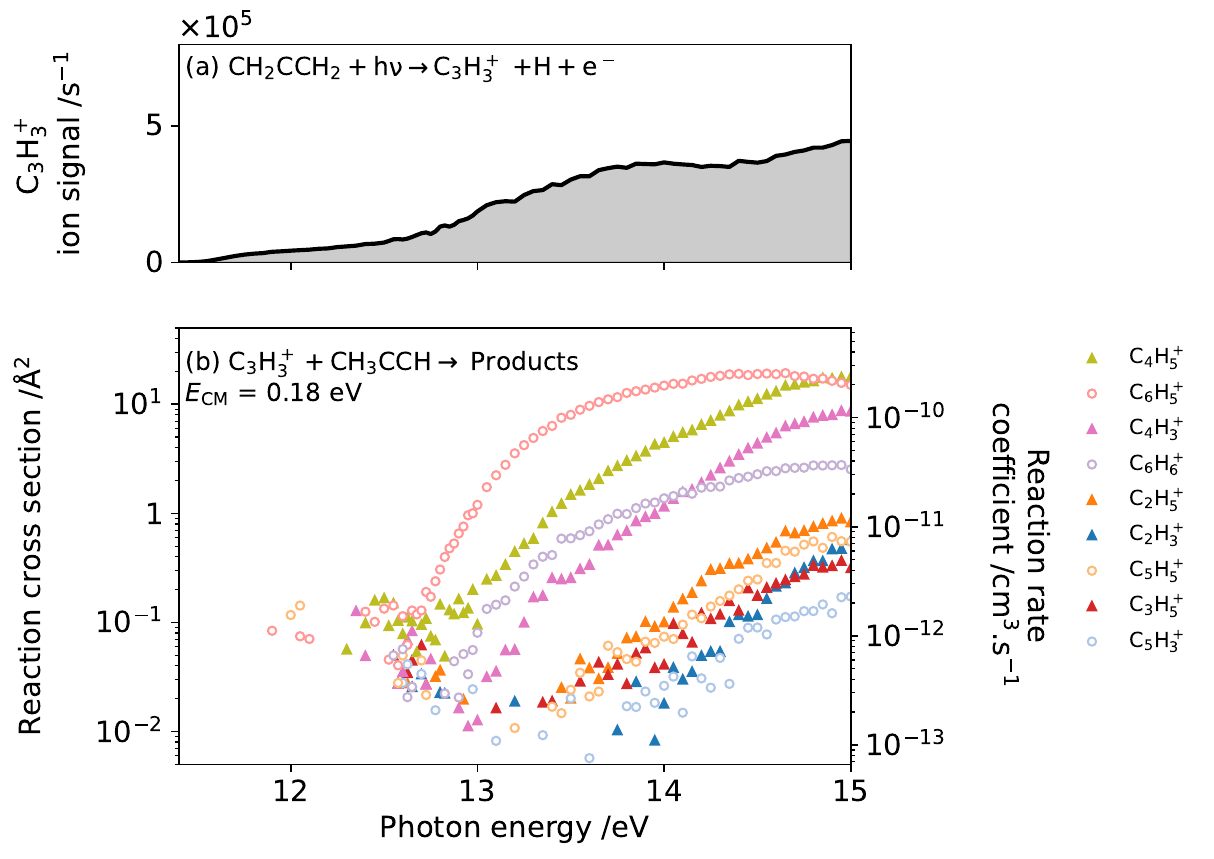}
\caption{Panels a displays the \ce{C3H3+} parent-ion signal from the dissociation photoionization of allene. Panels b shows reaction cross section (left axis) and rate coefficient (left axis) of all the reaction products of \ce{C3H3+ + C3H4}, as a function of the photon energy.} 
\label{SI_Allene_Propyne}
\end{figure}

\begin{figure}[!h]
\includegraphics[width=\linewidth]{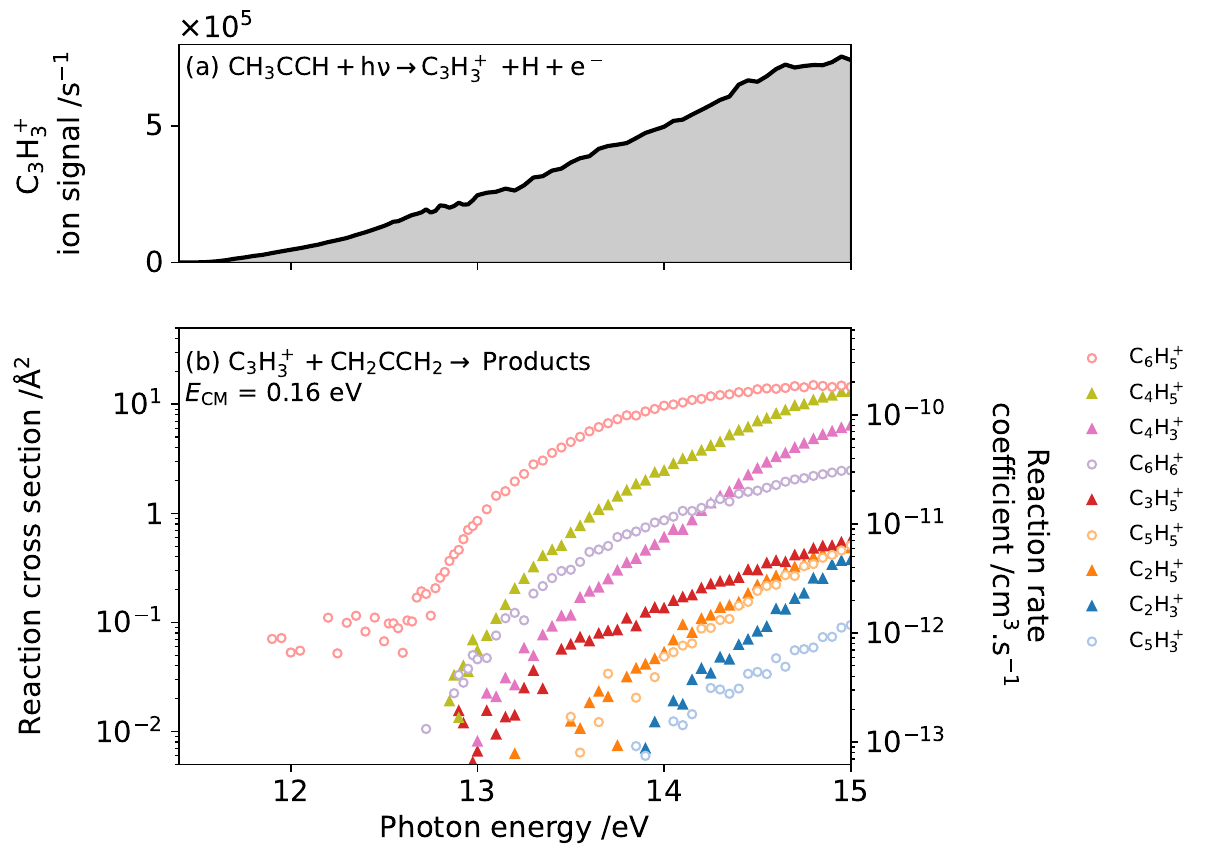}
\caption{Panels a displays the \ce{C3H3+} parent-ion signal from the dissociation photoionization of propyne. Panels b shows reaction cross section (left axis) and rate coefficient (left axis) of all the reaction products of \ce{C3H3+ + C3H4}, as a function of the photon energy.} 
\label{SI_Allene_Propyne}
\end{figure}

\begin{figure}[!h]
\includegraphics[width=\linewidth]{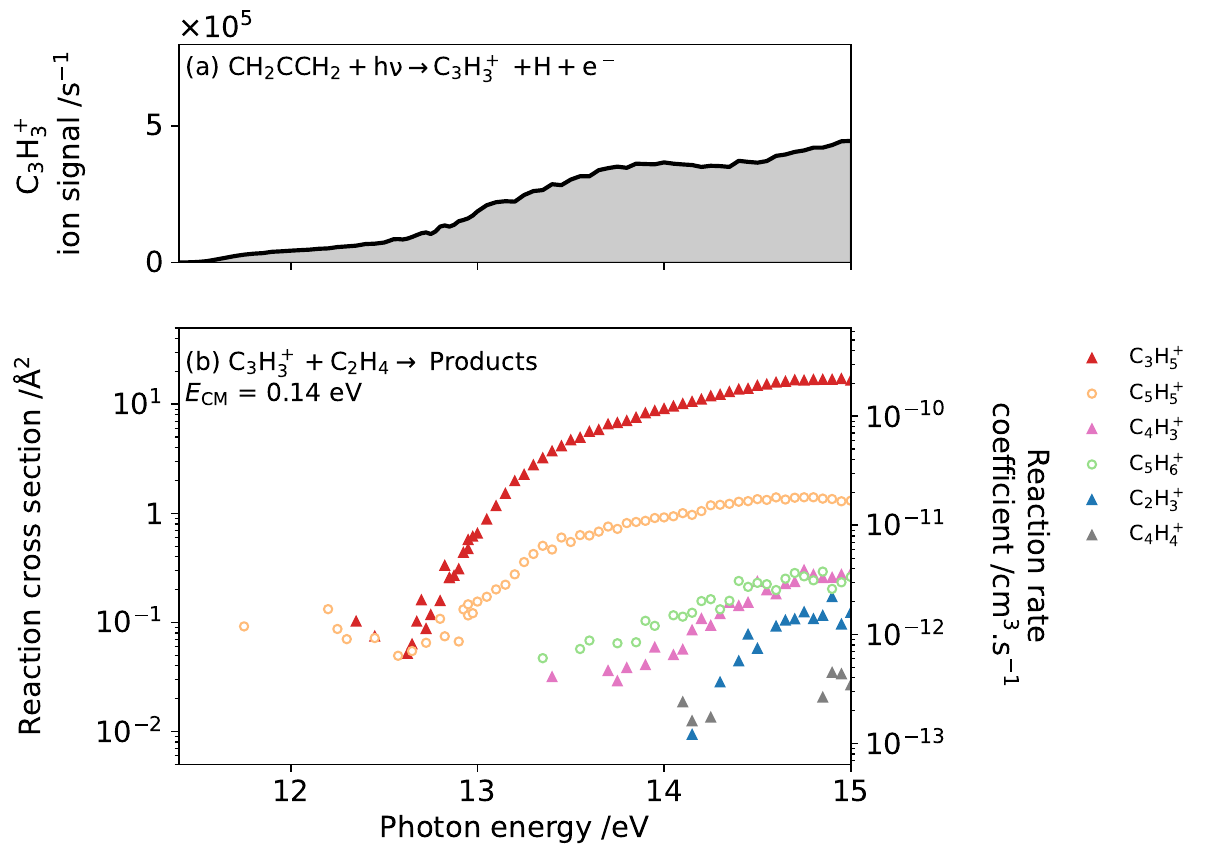}
\caption{Panels a displays the \ce{C3H3+} parent-ion signal from the dissociation photoionization of allene. Panels b shows reaction cross section (left axis) and rate coefficient (left axis) of all the reaction products of \ce{C3H3+ + C2H4}, as a function of the photon energy.} 
\label{SI_Allene_Propyne}
\end{figure}

\begin{figure}[!h]
\includegraphics[width=\linewidth]{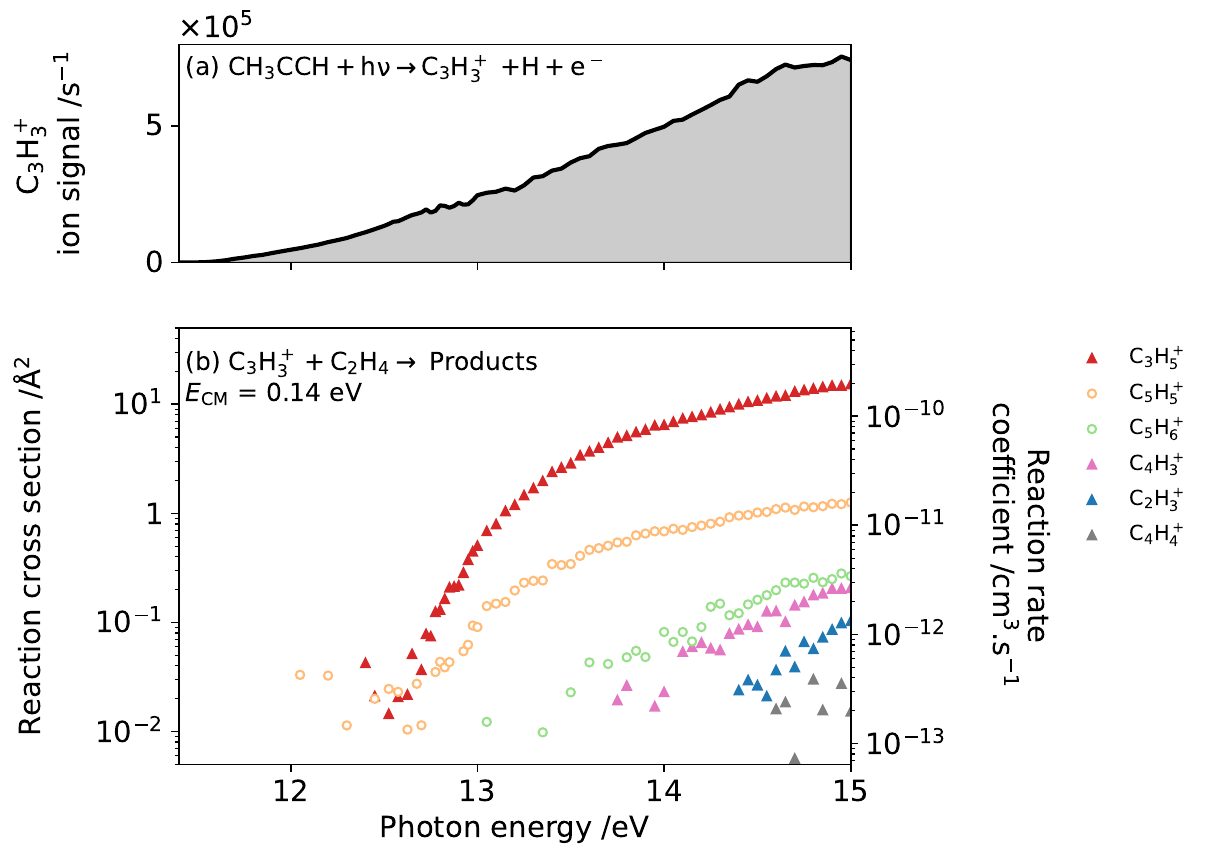}
\caption{Panels a displays the \ce{C3H3+} parent-ion signal from the dissociation photoionization of allene. Panels b shows reaction cross section (left axis) and rate coefficient (left axis) of all the reaction products of \ce{C3H3+ + C2H4}, as a function of the photon energy.} 
\label{SI_Allene_Propyne}
\end{figure}

\begin{figure}[!h]
\includegraphics[width=\linewidth]{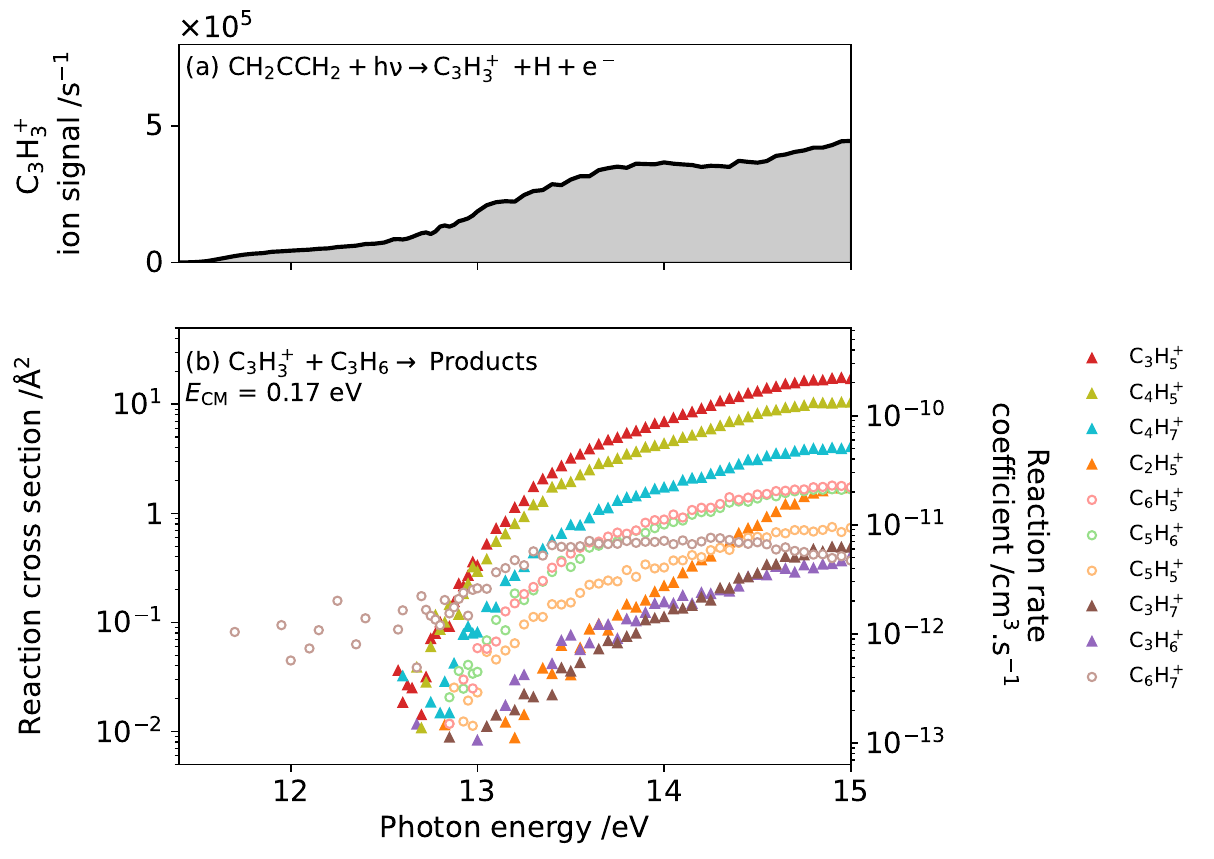}
\caption{Panels a displays the \ce{C3H3+} parent-ion signal from the dissociation photoionization of allene. Panels b shows reaction cross section (left axis) and rate coefficient (left axis) of all the reaction products of \ce{C3H3+ + C3H6}, as a function of the photon energy.} 
\label{SI_Allene_Propyne}
\end{figure}

\end{document}